\documentclass{ws-procs10x7}

\begin{document}

\title{--- Heavy Quark Phenomenology --- \\
$B\to \phi K^{(*)}$ CPV/Polarization, and Collider Implications}

\author{George Wei-Shu Hou}

\address{Department of Physics, National Taiwan
University, Taipei, Taiwan 106, R.O.C.\\E-mail:
wshou@phys.ntu.edu.tw}



\twocolumn[\maketitle\abstract{ %
The hint for BSM $CP$ violation in $B\to \phi K_S$ is now more
confused, but the $\phi K^*$ polarization anomaly seems real. We
present a picture based on a flavor-mixed, right-handed
``strange-beauty" squark $\widetilde{sb}_{1R}$, driven light by
the large $\tilde s_R$-$\tilde b_R$ squark flavor mixing, which
carries a unique new $CP$ phase. The $\widetilde{sb}_{1R}$ could
impact on $S_{\phi K_S}$ (or alternatively $S_{\eta^\prime K_S}$),
$B_s$ mixing, $\sin2\Phi_{B_s}$, $S_{K_S\pi^0\gamma}$ and other
$b\to s$ transitions, and can be searched for directly at the
Tevatron. Whether SM or BSM, a heuristic model is given where
transverse $\phi K^*$ polarization descends from the gluon
helicity of on-shell $b\to sg$. }]

\section{Introduction}

\subsection{Baryon \# Violating $\tau/c/b$ Decays?}\label{subsec:BNV}

Let me start by disclaiming a host of seemingly interesting baryon
number violating $\tau$, $D$ and $B$ decays.\cite{bnv} After
submission, we found that the stringent $p\to \pi\nu$ bound
implies ${\cal B}(\tau\to \bar p\pi^0) < 10^{-38}$, and slightly
weaker bounds apply to $\tau\to \bar \Lambda \pi^-$ by a weak
insertion. This agrees with an argument put forward by Marciano in
1995. Interestingly, Marciano's remark did not stop CLEO, in 1999,
from following the 1992 search by ARGUS for $\tau\to \bar p\pi^0$;
this year Belle searched for $\tau\to \bar \Lambda \pi^-$. Similar
arguments apply to $D$ and $B$ decays and Ref.~1 will be updated.

\subsection{Heavy Quarks and New Physics}\label{subsec:wpp}

The $b\leftrightarrow s$ transitions are arguably {\it the}
current frontier for New Physics (NP).
There is no sign of deviation in $b\leftrightarrow d$ phenomena
such as $B_d$ mixing and $\sin2\Phi_{B_d}$, but nagging
``discrepancies" in comparing $B\to K\pi$ and $\pi\pi$ transitions
have existed since 1999.

In 2003 Belle suggested $\sin2\Phi_{B_d\to \phi K_S}$ could be of
opposite sign to $\sin2\Phi_{B_d} \cong +0.73$. By adding an
equivalent amount of data, Belle\cite{sakai} now finds
$\sin2\Phi_{B_d\to \phi K_S}\simeq 0$, while BaBar's
result\cite{giorgi} is of the ``right" sign and only 1$\sigma$
below 0.73. However, BaBar now finds $\sin2\Phi_{B_d\to
\eta^\prime K_S}\simeq 0.27$ which seems low, although Belle finds
$0.65$ and is consistent with 0.73. But Belle and BaBar agree on
$\sin2\Phi_{B_d\to K_S \pi^0,\; K^+K^- K_S} \sim 0.3$, 0.5, which
offers some support for a lower $\sin2\Phi_{B_d\to \eta^\prime
K_S}$. Confirmation of a low $\phi K_S$ or $\eta^\prime K_S$ value
would indicate NP.

NP could also emerge in $b\to s\ell^+\ell^-$, a large $\Delta
m_{B_s}$, or $\sin2\Phi_{B_s} \neq 0$.
Another possibility would be\cite{chn} finding $b\to s\gamma_R$
(vs. $\gamma_L$ from SM). This can now be probed, thanks to
``$K_S$-tagging" technique developed by BaBar, by searching for
$\sin2\Phi_{B_d\to K_S\pi^0 \gamma} \neq 0$.
Thus, the study of $b\leftrightarrow s$ transitions offers a probe
of NP for decades to come.

\section{Light $\widetilde{sb}_{1R}$ and $\phi K_S \;
(\eta^\prime K_S)$ CPV}

The gist of our model\cite{chn} is a two-particle system
consisting of a right-handed, flavor-mixed $\widetilde{sb}_{1R}$
squark, of order 200 GeV hence rather light, and $\tilde g$ of
order 500--800 GeV. All other SUSY particles (except the
possibility of bino $\tilde\chi_1^0$) are at 1--2 TeV scale
because of low energy FCNC constraints.
The $\widetilde{sb}_{1R}$ squark brings in {\it a unique new CPV
phase $\sigma$} from $\tilde s_R$ and $\tilde b_R$ squark mixing,
as phase freedom has been used up in quark sector.

As right-handed flavor mixing is the least probed part of SM, a
light $\widetilde{sb}_{1R}$ squark could be just the right
particle to emerge from $b\leftrightarrow s$, {\it in our era of
heavy flavor factories}.

A light 500 GeV gluino seemed\cite{chn} needed with the 2003 Belle
result on $\phi K_S$. Indeed, one can see from Fig.~1 that
$\sin2\Phi_{B_d\to \phi K_S} < 0$ is rather drastic. We stopped at
500 GeV only for fear of FCNC bounds, and would resort to large
hadronic factors\cite{chn} had a negative result persisted.

\begin{figure}[t!]
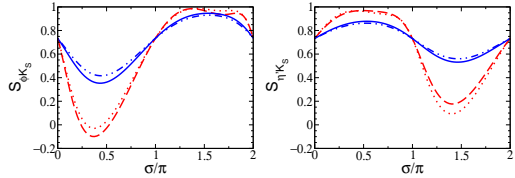

\includegraphics[width=1.3in,height=0.9in,angle=0]{sphik}
\includegraphics[width=1.3in,height=0.9in,angle=0]{setak}
\vspace{-1mm}
%
%
%
\caption{$S_{\phi K_S}$ and $S_{\eta^\prime K_S}$ vs $\sigma$ for
$m_{\widetilde{sb}_{1}}= 200$ GeV. Solid, dotdash (dash, dots)
lines are for common squark mass $\widetilde{m} =$ 2, 1 TeV,
$m_{\tilde g}=$ 0.8 (0.5) TeV.
}\label{fig:s}
\end{figure}

The case for $\sin2\Phi_{B_d\to \phi K_S} < 0$ has softened this
year, but new CPV results of Belle and BaBar are not in good
agreement in $\phi K_S$, $\eta^\prime K_S$ and $f_0(980) K_S$. We
ignore the latter since Belle and BaBar are of opposite sign and
the production dynamics for $f_0(980)$ is not known. We now see
too possibilities, due to the anticorrelation (in $\sigma$)
between\cite{chn,kou} $\sin2\Phi_{B_d\to \phi K_S}$ and
$\sin2\Phi_{B_d\to \eta^\prime K_S}$:

\noindent \ \ $\bullet$ {\it Scenario~1} --- If we take the
Belle/BaBar average value of $\sin2\Phi_{B_d\to \phi K_S} \sim
0.3$ and the new Belle result of $\sin2\Phi_{B_d\to \eta^\prime
K_S} \simeq 0.65$, then a more pleasant (in regards FCNC)
$m_{\tilde g}\sim$ 700 GeV with $\sigma \sim \pi/2$ would suffice.

\noindent \ \ $\bullet$ {\it Scenario~2} --- If the new BaBar
result of $\sin2\Phi_{\eta^\prime K_S} \sim 0.27$ reflects some
truth, let us combine with Belle and use $\sim 0.45$, but now take
$\sin2\Phi_{B_d\to \phi K_S} \sim \sin2\Phi_{B_d}$, then one could
have a different solution of $\sigma \sim 3\pi/2$, again with
$m_{\tilde g}\sim$ 700 GeV.

We emphasize these two new scenarios beyond published\cite{chn}
results in this proceedings.
Whether {\it Scenarios 1} or {\it 2}, the NP effect is still
large, hence the light $\widetilde{sb}_{1R}$ with its

\noindent \ \ 1) large effective $s$-$b$ mixing,

\noindent \ \ 2) a unique new CPV phase, and

\noindent \ \ 3) right-handed {\it strong} dynamics

\noindent involving the gluino, provides a natural setting. We
stress that the model is well motivated in its own
right:\cite{ch01} any underlying {\it approximate} Abelian flavor
symmetry would imply\cite{afs} large right-handed down flavor
mixings, {\it with $s_R$-$b_R$ mixing $\sim$ 1}. Invoking SUSY,
large flavor mixing repeats with squarks, and the strong dynamics
in face of FCNC constraints demand the need for 4 texture zeros in
the down-type quark mass matrix.\cite{ch01}

When no evidence emerged for NP effects involving
$b\leftrightarrow d$, we turned to decoupling $d$ flavor, which
decouples one from many L.E. constraints. We then found\cite{ach}
the interesting result that, not only near maximal $\tilde
s_R$-$\tilde b_R$ mixing could possibly drive
$\widetilde{sb}_{1R}$ light (with some tuning), the $RR$ mixing
effects are well hidden in $b\to s\gamma$: it is the induced $LR$
mixing effects that drive sensitivity near $\sigma \sim \pi$.

$B\to K^*\gamma$ and $b\to s\gamma$ illustrates FCNC constraints
and computational techniques. The effective $b\to s\gamma$
transition is induced by the r.h. (l.h.) $O_{11}^{(\prime)}$
operator with coefficient $c_{11}^{(\prime)}$, arising from
$t$-$W$ loop in SM ($\widetilde{sb}_{1R}$-$\tilde g$ loop). To get
$B\to K^*\gamma$ rate, one introduces the $B\to K^*$ form factor.
For 200 GeV $\widetilde{sb}_{1R}$ and $m_{\tilde g} = 500$ GeV,
$\pi/2 < \sigma < 3\pi/2$ is\cite{ach} ruled out by $b\to s\gamma$
data due to $LR$ mixing effects, but the whole range is allowed
for $m_{\tilde g} = 800$ GeV.
For hadronic modes such as $B\to \phi K_S$, the
$\widetilde{sb}_{1R}$ mainly\cite{ch01} feeds the color-dipole
$O_{12}^\prime$ operator (analogous to $O_{11}^\prime$), but now
one has large hadronic uncertainties.

So what are the implications for {\it Scenarios 1} or {\it 2} for
the near future? The situation is more relaxed with $m_{\tilde g}
\sim $ 700 GeV now allowed. $\Delta m_{B_s}$ would still be larger
than SM value, but would lie in the measurable 40--80 ps$^{-1}$
range. However, finding a large $\Delta m_{B_s}$, though
confirming NP, cannot tell apart the two scenarios; CPV
measurements are needed. Besides confirmation of either
$\sin2\Phi_{B_d\to \phi K_S}$ or $\sin2\Phi_{B_d\to \eta^\prime
K_S}$ being low, with $K_S\pi^0$ (model dependence very similar to
$\eta^\prime K_S$) modes as crosscheck but providing no further
insight (due to hadronic uncertainties\cite{chn}), one needs some
additional CPV measurement to resolve $\sigma$.

\begin{figure}[t!]
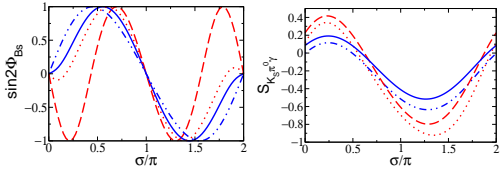

\includegraphics[width=1.27in,height=0.86in,angle=0]{sin2Phi}
\hspace{-1mm}
\includegraphics[width=1.27in,height=0.83in,angle=0]{mix_f}
\vspace{-1mm}
%
%
\caption{$\sin2\Phi_{B_s}$ and $S_{K_S\pi^0\gamma}$ vs $\sigma$,
with parameters and notation as in Fig.~1.}\label{fig:cpv}
\end{figure}

We plot $\sin2\Phi_{B_s}$ and $\sin2\Phi_{B_d\to K_S\pi^0\gamma}$
(where $K_S\pi^0$ forms a $K^{*0}$) in Fig.~2.
The clean measurables $\Delta m_{B_s}$ and $\sin2\Phi_{B_s}$ have
little hadronic uncertainties. They probe the
$\widetilde{sb}_{1R}$-$\tilde g$ box hence are very sensitive to
the masses involved, so provide information on these parameters.
Furthermore, finding $\sin2\Phi_{B_s} \sim +1$ or $-1$ would
provide striking confirmation of {\it Scenarios 1} or {\it 2},
i.e. either a low $\sin2\Phi_{B_d\to \phi K_S}$ or
$\sin2\Phi_{B_d\to \eta^\prime K_S}$, which should become clear by
2007.

Our two-particle NP model has 3 parameters,
$m_{\widetilde{sb}_{1R}}$, $m_{\tilde g}$ and $\sigma$ (one is
less sensitive to a 4th, $m_{\widetilde{sb}_{2R}} \cong \sqrt 2\,
\widetilde m$), so a third clean measurable is needed. Parameters
such as $\sin2\Phi_{B_d\to \phi K_S}$ or $\sin2\Phi_{B_d\to
\eta^\prime K_S}$ have murky hadronic uncertainties. Of course, a
200 GeV $\widetilde{sb}_{1R}$ squark can be searched for directly
at the Tevatron, which we turn to later. Surprisingly, an
additional clean CPV measurable exists: $B_d\to K^{*0}\gamma \to
K_s\pi^0\gamma$.

The photon in $B \to K^*\gamma$ would be dominantly r.h. within
SM. The strong r.h. $\widetilde{sb}_{1R}$-$\tilde g$ dynamics
would produce a l.h. photon, $\propto c_{11}^\prime$ at amplitude
level. This allows for $t$-dep. CPV interference\cite{ags} between
$B^0$ and $\bar B^0$ decays, suppressed by $m_s^2/m_b^2$ in SM.
The formula for $\sin2\Phi_{B_d\to K_S\pi^0\gamma}$ is simply
\begin{equation}
\frac{2\vert c_{11}c^\prime_{11}\vert}{\vert c_{11}\vert^2 + \vert
c^\prime_{11}\vert^2} \,
\sin\left(2\Phi_{B_d}-\varphi_{11}-\varphi_{11}^\prime\right),
\end{equation}
where one sees that there are no hadronic uncertainties, with
$\vert c_{11}^\prime \vert$ and $\phi^\prime \equiv \arg
c_{11}^\prime$ all computable in our model ($c_{11} \cong
c_{11}^{\rm SM}$).

We see from Fig.~2 that $\sin2\Phi_{B_d\to K_S\pi^0\gamma}$ can
not only crosscheck the sign of $\sin2\Phi_{B_s}$, the strength
also offers very valuable information.
In {\it Scenario 1}, $\sin2\Phi_{B_d\to K_S\pi^0\gamma} \sim +0.1$
would be much harder to measure than $\sin2\Phi_{B_d\to
K_S\pi^0\gamma} \sim -0.6$ in {\it Scenario 2}. In this sense, we
hope BaBar is right, that it is $\sin2\Phi_{B_d\to \eta^\prime
K_S}$ (together with a few other $PP$ modes) that is lower than
$\sin2\Phi_{B_d}$. After all, Belle's measurement of
$\sin2\Phi_{B_d\to \phi K_S}$ utilizing their upgraded SVD2
silicon detector (2004 data) is consistent with $\sin2\Phi_{B_d}$.
In any case, $\sin2\Phi_{B_d\to K_S\pi^0\gamma}$ offers a third
clean measurement to determine our model parameters from
$b\leftrightarrow s$ studies.

\section{$\phi K^*$ Polarization Puzzle}

Belle and BaBar now agree\cite{sakai} on the longitudinal fraction
$f_0 \cong 0.5$ in $B\to \phi K^*$, and transverse $f_\perp \simeq
f_\parallel \cong 0.25$. This confirms the $\phi K^*$ polarization
puzzle since last year, against the factorization expectation of
$f_0 = 1 - {\cal O}(1/m_b^2)$; $f_\perp/f_\parallel = 1 + {\cal
O}(1/m_b)$ now seems respected.
Since $f_0 = 1$ holds in tree dominant $\rho^+\rho^-$ and
$\rho^+\rho^0$ modes, and seemingly for $\rho^0 K^{*+}$, the quest
this year has been whether the naively pure $b\to s\bar dd$
penguin $\rho^+ K^{*0}$ mode emulates $\phi K^*$ ($b\to s\bar
ss$).

Indeed, Belle finds $f_0 \simeq 0.50$ for $\rho^+ K^{*0}$, while
BaBar finds $0.79$. But they disagree on rate ($\simeq 6.6\times
10^{-6}$ vs $17.0\times 10^{-6}$), so we can only conclude
$f_0(\rho^+ K^{*0}) < 1$.

\begin{figure}[b!]
\begin{center}
\vspace{-2mm}
\includegraphics[width=1.4in,height=0.73in,angle=0]{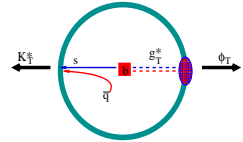}
\vspace{-2mm}
\end{center}
%
%
\caption{Heuristic picture for $\phi_T$
emission.}\label{fig:trans}
\end{figure}

We proposed\cite{hn} a heuristic but {\it ad hoc} picture (Fig.~3)
to account for the $\phi K^*$ polarization anomaly.
Our observation starts with $B\to K^*\gamma$. The $K^*$ is purely
transverse here because the photon is transverse. Inclusive $b\to
s\gamma$ has rate $\sim 3\times 10^{-4}$, and has a gluonic cousin
of $b\to sg$, which has rate $\sim {\rm few} \times 10^{-3}$
(0.1\% in\cite{bsg} SM). Clearly {\it this on-shell gluon is
transverse}, but unlike the photon which immediately escapes, the
gluon ``fragments" into a ``jet". $f_0 \cong 0.5$ means transverse
$B\to K_T^* \phi_T \sim 5\times 10^{-6}$. While $B\to K^*\gamma$
is about 10\% of $b\to s\gamma$, if only one per mille of $b\to
sg$ gives $B\to K_T^* \phi_T$, the $\phi K^*$ anomaly is accounted
for.
%
%
The ``gluon remnant" (left at hadron surface) and the recoil $s$
and spectator $\bar q$ quark form a color singlet. We introduced a
new hadronic parameter for $B\to \phi_T K^{*}_T$, and as
corollary, $\omega K^*$.

Our gluon fragmentation picture applies to both SM and our
$\widetilde{sb}_{1R}$ model. For the latter, with softening of the
$\phi K_S$ anomaly (alternatively $\eta^\prime K_S$), we expect
$T$ violating triple products to be consistent with present data,
while CPV results are not yet reported.

We note that the singlet nature of the gluon implies that this
process does not affect charged vector meson or $\rho^0$. If
$f_0(\rho^+ K^{*0}) = 0.5$ as suggested by Belle, we'll probably
accept defeat. However, if the value is 0.8, we suggest
disentangling isospin components, since transverse $B\to \phi_T
K^{*}_T$ and $\omega_T K^*_T$ can rescatter into $I = 1/2$ part of
$\rho K^*$.

\section{Collider Search for Light $\widetilde{sb}_{1R}$} 

Even though $B$ studies can fully determine
$m_{\widetilde{sb}_{1R}}$, $m_{\tilde g}$ and the new CP phase
$\sigma$, nothing beats direct observation of a new particle. In
our TeV scale SUSY model, the $\widetilde{sb}_{1R}$ is {\it driven
light by large flavor mixing}, which is unusual. We have
shown\cite{ach} that a bino $\widetilde \chi_1^0$ lighter than
even 100 GeV is allowed by $b\to s\gamma$. But unlike the
$\widetilde{sb}_{1R}$, we have no argument for why $\widetilde
\chi_1^0$ (nor the gluino itself) is light.

We stress that standard $\tilde b$ squark limit is diluted by dual
$s$-$b$ flavor of the $\widetilde{sb}_{1R}$, since
$\widetilde{sb}_{1R} \to b \widetilde \chi_1^0$, $s \widetilde
\chi_1^0$ (or gravitino $\widetilde G^0$). This should be kept in
mind for direct search.
A 200 GeV squark certainly can be uncovered by the Tevatron, and a
few hundred raw events per few fb$^{-1}$ luminosity is expected,
with $q\bar q$ process for standard $\tilde b$ squark production
dominating over $gg$. Discovery is not a problem, but the flavor
mixing angle $\sin^2\theta_m$ controls the $b$ fraction, hence
double $b$-tagged events are diminished. Nevertheless good
$b$-tagging is crucial, and the single vs. double tag ratio
contain information on $\sin^2\theta_m$.

\begin{table}[t!]
\footnotesize \caption{\label{tab:f} Tevatron cross sections (fb)
for $\widetilde{sb}_{1R}$. }
\vspace{-0.07cm} \centerline{
\begin{tabular}{|c|ccc|}
\hline
 Mass
          & 0 $b$-tag
          & 1 $b$-tag
          & 2 $b$-tag
          \\
\hline 150
          & 283
          & 243
          & 51
          \\
200
          & 68
          & 61
          & 14
          \\
250
          & 16
          & 15
          & 3.3
          \\
300
          & 4.0
          & 3.7
          & 0.83
          \\
350
          & 1.0
          & 0.93
          & 0.21
          \\
\hline
\end{tabular}
}
\end{table}

Let us take maximal mixing, i.e. $\sin^2\theta_m = 0.5$ as
reference. Table 1 gives cross sections at the Tevatron, where we
see the $\widetilde{sb}_{1R}$ can be discovered ($> 10$ events
with low background) up to 300 GeV.

\section*{Acknowledgments}
I thank C.K. Chua, K. Cheung, M. Nagashima and A. Soddu for
collaboration.

\end{document}